\documentclass[aps,pra,twocolumn,showpacs,amsmath,superscriptaddress]{revtex4-1}

\usepackage{amsmath}
\usepackage{amsfonts}
\usepackage{amssymb}
\usepackage{mathrsfs}
\usepackage{graphicx}
\usepackage[justification=raggedright]{caption}
\usepackage{color}
\usepackage{verbatim}
\usepackage{floatrow}
\usepackage{bm}
\usepackage{lipsum}
\usepackage{tabularx}
\usepackage[colorlinks,linkcolor=blue,citecolor=blue,urlcolor=blue]{hyperref}
\usepackage[hang,center, nooneline, Large]{subfigure}

\usepackage[usenames,dvipsnames]{xcolor}
\usepackage{soul}
\newcommand{\new}[1]{\textcolor{blue}{#1}}

\begin{document}
	\title{Trapping and Sympathetic Cooling of Boron Ions}
	\author{Ren\'{e} Rugango}
	 \affiliation{Schools of Chemistry and Biochemistry, Georgia Institute of Technology, Atlanta, GA 30332}
	\author{Mudit Sinhal}
	 \affiliation{Schools of Chemistry and Biochemistry, Georgia Institute of Technology, Atlanta, GA 30332}
        \affiliation{Department of Physics and Meteorology, Indian Institute of Technology , Kharagpur, India 721302 } 
	\author{Gang Shu}
	 \affiliation{Schools of Chemistry and Biochemistry, Georgia Institute of Technology, Atlanta, GA 30332}
	\author{Kenneth R. Brown}
	\email{kenbrown@gatech.edu}
         \affiliation{Schools of Chemistry and Biochemistry, Georgia Institute of Technology, Atlanta, GA 30332}
          \affiliation{Schools of Computational Science and Engineering, and Physics, Georgia Institute of Technology, Atlanta, GA 30332}
	\date{\today}

\begin{abstract}
We demonstrate the trapping and sympathetic cooling of B$^{+}$ ions in a Coulomb crystal of laser-cooled Ca$^{+}$, We non-destructively confirm the presence of the both B$^+$ isotopes by resonant excitation of the secular motion. The B$^{+}$ ions are loaded by ablation of boron and the secular excitation spectrum also reveals features consistent with ions of the form B$_{n}^{+}$. 
\end{abstract}

\maketitle

\section{Introduction}

Boron atomic ions, with their Be -like electronic structure, are a potential candidate for a high precision atomic clock \cite{ludlow2015} with a predicted blackbody uncertainty close to 1 $\times$ 10$^{-18}$ at room temperature \cite{Safronova2012}. Directly laser-cooling B$^+$ is challenging due to the required short wavelengths to drive the transitions.  A B$^+$ ion clock could be built using another laser-cooled atomic ion for both cooling and readout following the success of the Al$^+$ clock \cite{wineland, ChouPRL2010, Rosenband2008} . The four-electron closed-shell electronic structure of B$^{+}$ ions also makes them good candidates as sensors for the analysis of astrophysical plasmas \cite{Muller2010}. Boron clusters and boron-hydrides are also of wide interest due to the importance of B in materials fabrication \cite{Takayuki2005, Mukerjee2016}. B$_{n}^{+}$  ions have been previously loaded into ion traps using ablation of a boron target \cite{hanley1987, Marianne1997, Ruatta1989} and we use this method to load B$_n$ ions into a trap containing laser-cooled Ca$^+$ ions.

Molecular and atomic ions sympathetically cooled by laser-cooled atomic ions have proven to be a useful platform for studying molecular reactions and spectroscopy. Large ion crystals have provided insight into molecular spectra \cite{RugangoCPC2016,BresselPRL2012, RellegertNature2013}, while small crystals have enabled the translational cooling of molecular ions to microkelvin temperatures \cite{rugango2015, wan2015}, the non-destructive spectroscopy of vibronic transitions  \cite{wolf2015}, and the implementation of quantum-logic based atomic clocks \cite{wineland, ChouPRL2010}. High precision spectroscopy of molecular ions in Coulomb crystals has direct applications to the measurement of the electron electric dipole moment and the time variation of fundamental constants \cite{KajitaJPhysB2011, BresselPRL2012}.

Here we present the sympathetic cooling and trapping of B$^+$, B$_2^+$, and B$_3^+$ by Ca$^+$ ions. The B$_n^+$ ions are produced by ablating a target of pure elemental B and the ions are identified by modulating the fluorescence through resonant excitation of the motion. Peaks compatible with boron hydride ions are also occasionally observed, but not reliably produced.  In the next section we describe our production and characterization techniques. In section \ref{results} we discuss our observations and the relative occurrence of different ion species. Finally we conclude with some remarks on future experiments.

\section{Experimental Methods}

Details on the setup used in this experiment can be found  in Ref. \cite{RugangoCPC2016} and \cite{goeders}. The ions are trapped using a linear Paul trap with eleven segments housed in a spherical octagon vaccum chamber (Kimball Physics MCF800-SphOct-G2C8). Low DC voltages (0 - 10V) applied to all eleven pairs of DC electrodes weakly hold the ions in the axial direction. We typically use a RF voltage of 198 V oscillating at 19.35 MHz to confine ions radially. At this frequency, low Matthieu $q$ values for both stable $^{11}$B$^{+}$ ($q_{B^+}$ = 0.23)  and $^{40}$Ca$^{+}$ ($q_{Ca^+}$ = 0.064) ions are achieved.  These trap parameters result in a radial secular frequency of 0.44 MHz for Ca$^{+}$. The trap voltage is varied  between 148 V and 198 V corresponding to secular frequencies of 0.33 to 0.44 MHz.  The Mathieu $a$ parameter is estimated to be \textless 0.001 for Ca$^+$. Ca$^+$ ions are Doppler cooled axially and radially using 397 nm and 866 nm lasers.

The Ca$^{+}$ ions are loaded by photoionizing neutral, thermally-evaporated Ca with 423 nm and 379 nm lasers or by ablating a Ca target using a 355 nm nanosecond YAG laser  (Continuum minilite II) \cite{Zimmermann2012}. The boron ions are subsequently trapped by ablating the target mounted next to the Ca target. We monitor the power of the YAG to avoid too many hot boron ions that would destabilize the already trapped Ca$^{+}$ crystal.  We maintain the power around 6 mW corresponding to 6 mJ/pulse for fast loading of boron ions and to prevent loss of the Ca$^+$ crystal.  A CCD camera (Princeton Instruments Cascade 1K) and a photomultiplier tube (Hamamatsu H7360-02) are used to image and count the fluorescence of the Ca$^{+}$ ions. The Ca$^{+}$ ions sympathetically cool the B$^{+}$ ions and facilitate their trapping. 

The dark  boron ions are detected using the motional resonance coupling method \cite {krems, Baba2001} by applying an oscillating voltage on the top center electrode of the trap and observing changes in the fluorescence as function of the oscillation frequency. The radial motion of a single ion scales as the charge to mass ratio, $Q/m$, and in a crystal ions of different mass form bands near the bare resonance frequency. This method has been used to observe many sympathetically-cooled species such as HD$^{+}$ \cite{BlythePRL2005}, BeH$^{+}$ \cite{Roth2006}, C$_{60}^{+}$ \cite{SchuesslerPRA2006}, C$_{6}$H$_{5}$NH$_{2}^{+}$ \cite{DrewsenPRA2008}, Zn$^{+}$, $^{44}$Ca$^{+}$, and  Ga$^{+}$ \cite{Yoshiki2001}. Typical shifts in terms of $\Delta(Q/M)$ of up to 25 $\%$ due to both Coulombic interaction and around 10 $\%$ due to  stray-fields mixing modes \cite{Roth2007} have been observed. To minimize these shifts, we compensate for stray fields using  the time-average average position and the fluorescence modulation techniques \cite{berkeland1998, Pyka2014} and weakly confine the ions in the axial direction to reduce the strength of the Coulomb interaction. In addition, compensation for excess electric fields reduces heating due to driven RF motion, known as micromotion.

\begin{figure}
\includegraphics[width=9cm]{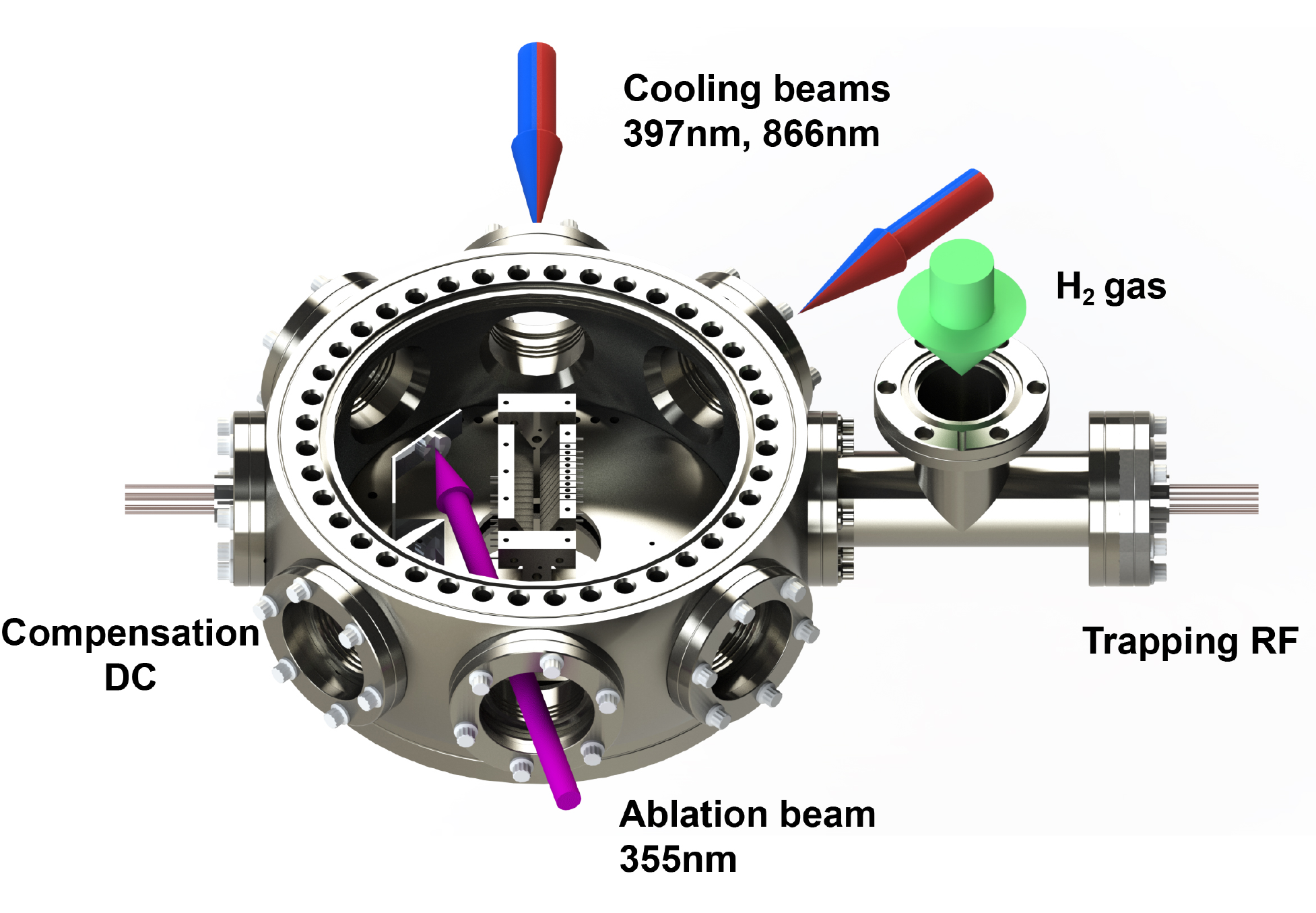}	
\caption{The experimental setup including the main vacuum chamber and the trap is shown. The The 355 nm YAG minilite II is focused on B target is mounted right next to the Ca target about 40 cm from the trap center. The cooling beams (397 nm and 866 nm)used to Doppler cool the ions enter the trap axially and radially. }
\label{fig:Apparatus}
\end{figure}

\section{Results and Discussion} \label{results}

\begin{figure*}
\centering
\includegraphics[width=18cm]{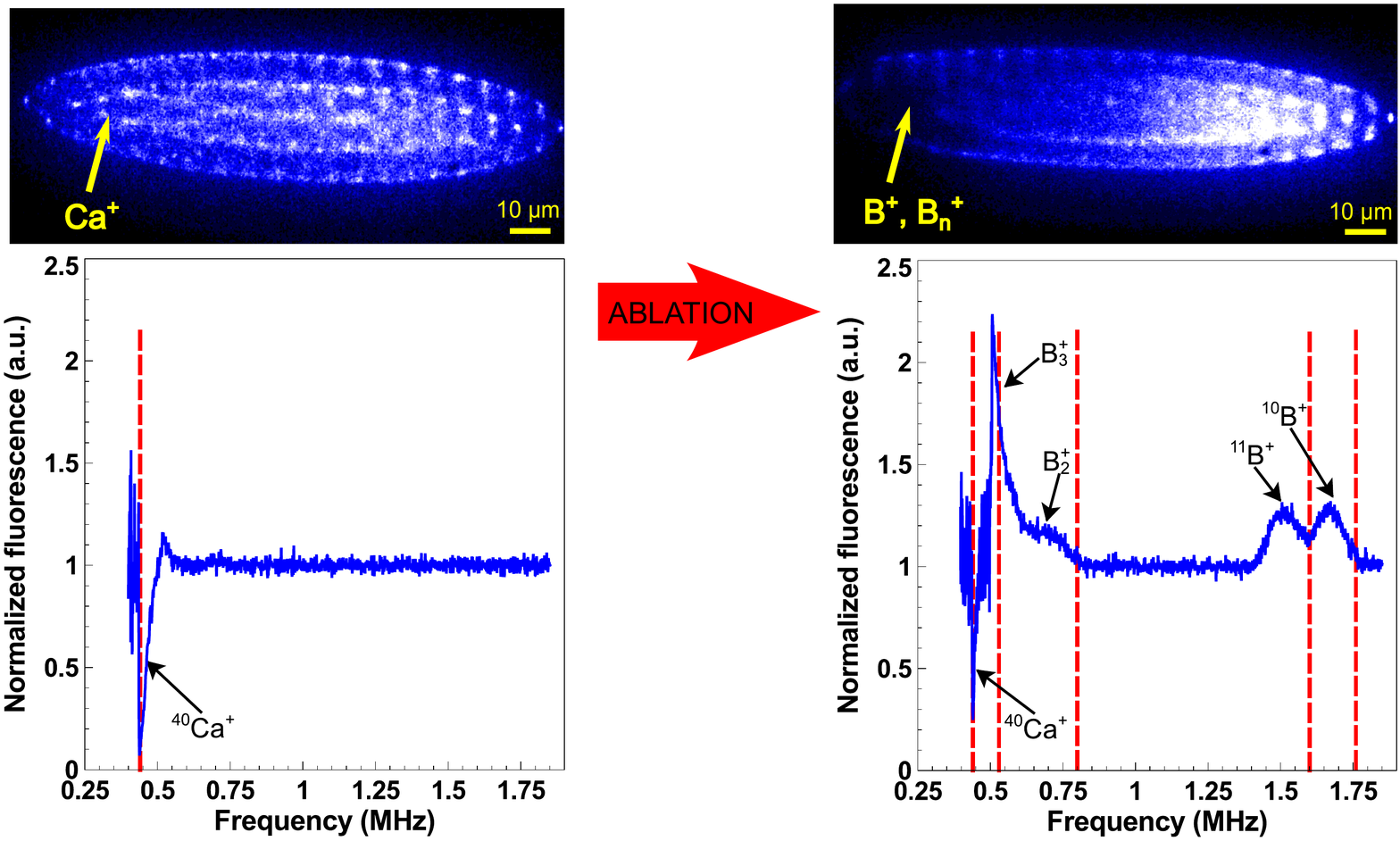}
\caption{Ion crystals containing about 200 Ca$^{+}$ ions with their corresponding secular excitation scans before and after ablation of the boron target. The boron species are located towards the center of the crystal due to a lower mass-to-charge ratio, and the secular frequencies for different ions are inversely proportional to their $m/q$. The vertical bar are the predicted secular frequencies for each ion. One can observe that the peak positions are within 10\% of their predicted values.  In this figure, $\omega_{Ca^+}$ is set to 0.44 MHz and $q_{B{^+}}$=0.23.}
\label{fig:scans}
\end{figure*}

Our reference for mass determination is a single $^{40}$Ca$^{+}$ ion with a secular frequency set between 0.33 MHz and 0.44 MHz. Initial ablation of Ca$^+$ yields a secular excitation spectrum with a single band at the expected location with a width of 3 kHz  for a typical applied excitation voltage of 0.35 V.  Ablation of the B target then introduces extra peaks into the spectrum. An example of motional frequency spectra before and after ablating the boron target are shown in Fig. \ref{fig:scans}.  We consistently see two weak peaks and a strong peak with a shoulder. These features are consistent with $^{10}$B$^+$, $^{11}$B$^+$, B$^+_2$ , and B$^+_3$.    The secular frequencies that we measure after trapping boron species are within 10 \% of the expected values.  Since the frequency spacing scales as $\Delta M/M_{1}M_{2}$ for singly charged ions, the resolution is better for lighter ions and prevents us from distinguishing isotopically distinct molecular ions.

\begin{table*}
	\centering
	\begin{center}
		\begin{tabularx}{0.8\textwidth}{c c c c c}
			\hline\\

			 Assignment & Predicted position (MHz) & Actual position (MHz) & Mass range (amu) & Times observed\\
\\
			\hline\\
			$^{10}$B$^{+}$ & 1.48 & 1.40 & 10.2 - 11.0 & 6\\ 
			$^{11}$B$^{+}$ & 1.34 & 1.28 & 11.0 - 12.3 & 10\\ 
			B$^{+}_{2}$ & 0.67 & 0.62 & 21.1 - 27.9 & 8\\ 
			B$^{+}_{3}$ & 0.45 & 0.45 & 30.8 - 37.9 & 10\\ 
			Ca$^{+}$ & 0.37 & 0.37 & 38.0 - 42.2 & 10\\
			Unknown & - & 1.05 & 13.4 - 14.9 & 1\\
			\hline
		\end{tabularx}
		
		\caption{A table showing the trapping of different ion species by ablation and their occurrence for 10 consecutive loading events. The last column shows how many times a species is observed out of ten motional frequency scans on different crystals. The peak observed between 13.6 and 14.6 amu has not been identified but could correspond to BH$_{2}^{+}$. The $^{10}$B$^{+}$ isotope appears less frequently and its peak is usually shorter than that of $^{11}$B$^{+}$. The shifts from the predicted peak positions are due to Coulombic interactions between the ions.}
\label{histogram}
	\end{center}
\end{table*}

The mass of the trapped ions determines their stability in the trap as well as sympathetic cooling efficiency. Consequently, those with masses similar to that of $^{40}$Ca$^{+}$ are the most stable. Fig \ref{fig:scans} shows typical secular excitation scans before and after loading boron species. We observe a strong signal from B$_{3}^{+}$  due to its mass similarity with Ca$^{+}$. Furthermore, previous studies on B$_{3}^{+}$  have also shown it to be a relatively stable cluster compared to its lighter counterpart (B$_{2}^{+}$) \cite{hanley1987}. We are also occasionally able to detect both isotopes of boron as shown in Fig. \ref{fig:scans}, but we are unable to resolve or detect the clusters based on both isotopes.

\begin{figure}
\centering
\includegraphics[width=\columnwidth]{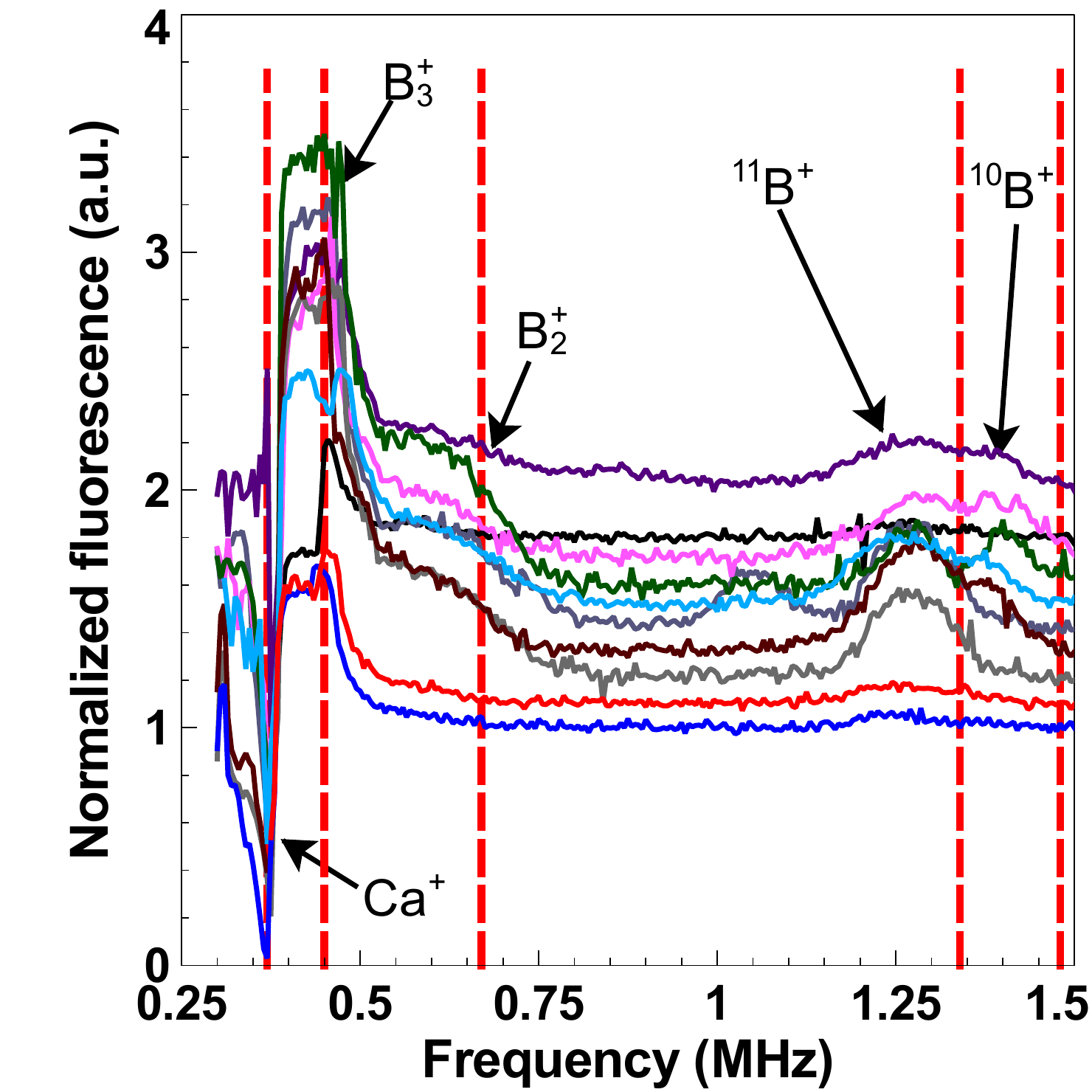}
\caption{Secular excitation scans are depicted for 10 consecutive experiments.  For each experiment, a new Ca$^+$ crystals is loaded and the B target is ablated 10 times.  \new{On the y-axis the plots are offset from each other by 0.1 units. The features corresponding to B$_3^+$ and $^{11}$B$^+$ are observed in every scan. The difference in peak heights between different scans stems from the fact that ablation does not allow control over the  amount of dark ions loaded in each crystal.}}
\label{fig:multi-scan}
\end{figure}

To determine the reliability of the ablation method for producing different B$_n^+$ species, we performed 10 consecutive experiments  where a new Ca$^+$  ion crystal of 200-300 ions was loaded before the boron target was ablated. The resulting mass excitation spectrum are shown in Fig. \ref{fig:multi-scan} and the summary of detected species is presented in Table. \ref{histogram}. We have found the atomic and molecular boron ions are consistently loaded.  In addition an unassigned peak that could correspond to BH$_{2}^{+}$ \cite{jeff1992, dryza2008} was observed. The $^{10}$B$^{+}$ isotope is loaded less frequently but at a rate higher than its natural abundance. Such isotope enrichment in ablation plumes has been previously noted for boron targets and other species like titanium, zinc, copper, and gallium \cite{Pronko1999, Manoravi2003}. 

\begin{figure}
\centering
\subfigure[hang, center][$q$ = 0.23]{\label{fig:a}\includegraphics[width=80mm]{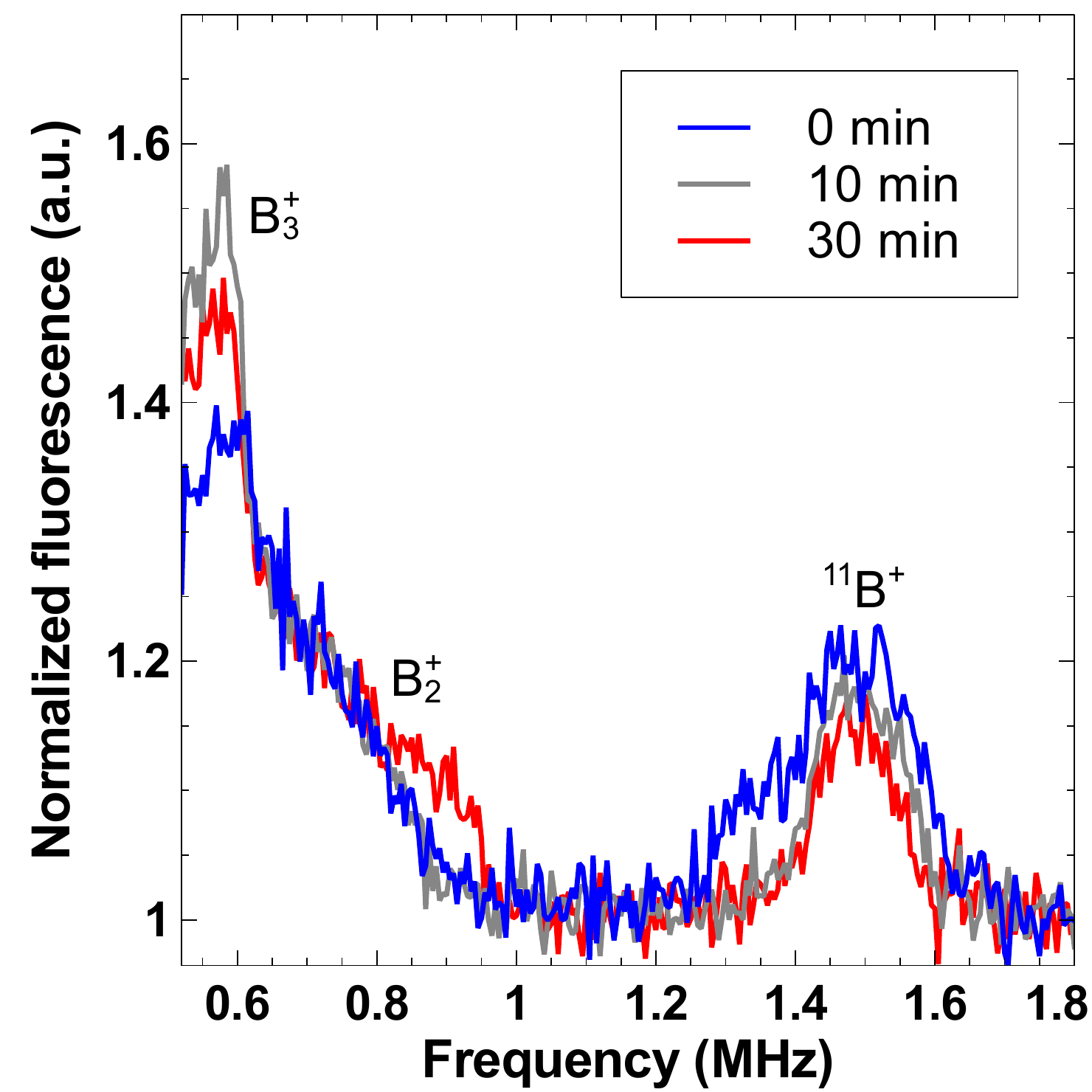}}
\subfigure[$q$ = 0.18]{\label{fig:b}\includegraphics[width=80mm]{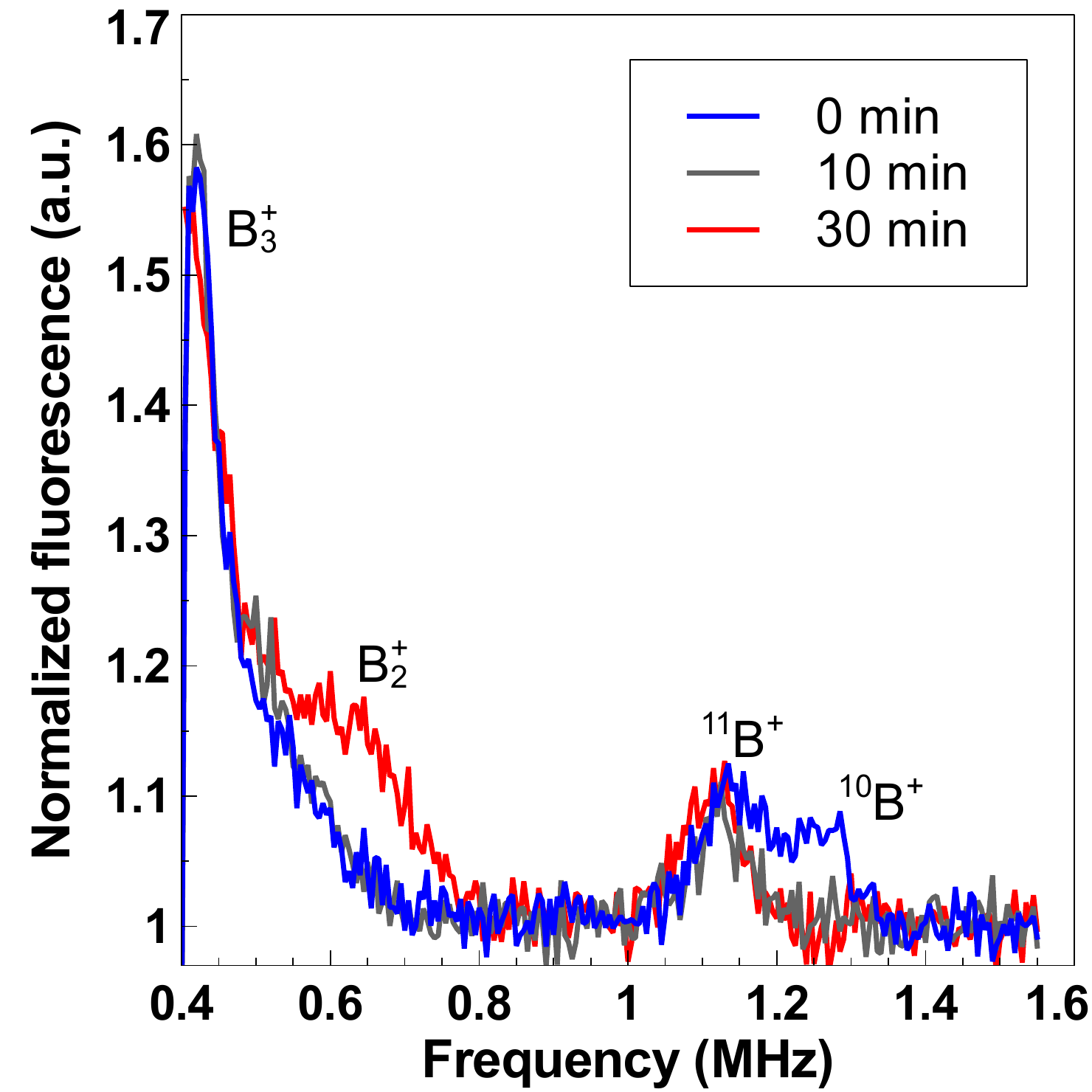}}
\caption{The secular excitation spectrum can change with trapping time. As seen in a) and b), the shoulder feature that we associate with B$_2^+$ can appear a long time after ablation.  This could be either due to dissociation of B$_4^+$ whose presence is hidden by the Ca$^+$ line or charge exchange with background gas.  The shoulder off the $^{11}$B$^+$ peak in a) is compatible with BH$^+$ but the feature has a short lifetime and is rare. The features corresponding to B$^+$ and B$_3^+$ are observed to be relatively stable over thirty minutes.}
\label{B2lifetime}
\end{figure}

We have observed that the detected peaks change dynamically with time.  First, we have observed that the $^{11}$B$^+$ ions are stable at both $q_{B^{+}}=0.23$ and $q_{B^{+}}=0.18$. Shorter lifetimes for $^{11}$B$^+$ ions are occasionally observed when collisions with background gas surge. A titanium sublimation pump is used to lower the rate of those collisions. Second, although we normally observe the shoulder feature that we associate with  B$_2^+$ directly after ablation, it occasionally grows in at longer times (Fig. \ref{B2lifetime}).  We have not seen a correlation between the growth of this shoulder and loss of signal in other peaks. One possible explanation could include dissociation or ionization of B$_4^+$ ions that are hidden by the Ca$^+$ transition or charge exchange of Ca$^+$ with background gases. A complete understanding of the process requires improved $Q/M$ resolution in this region. Finally, we also occasionally see a broadening of the B$^+$ peak that could represent BH$^+$ (Fig. \ref{B2lifetime}). Out of  hundreds of  loading events, we have only observed such a peak twice. Increasing H$_2$ pressure up to 1$\times$10$^{-7}$ Torr did not yield additional BH$^+$.

 BH$^{+}$ has an internal structure that is potentially suitable for Doppler cooling with a few lasers \cite{NguyenNJP2011}, but its gas phase production has been limited by the high melting point of elemental B \cite{chemistryhandbook} and the high activation energies needed for reactions \cite{ottinger1981}. Also, unlike other metal hydrides ions  which can be produced from photoactivated reactions between laser cooled atomic ions and H$_{2}$ \cite {kimura, Roth2006, MolhavePRA2000}, B$^{+}$ has not been laser cooled yet. Previous methods of BH$^{+}$ production in the gas phase relied on hazardous precursors such as B$_{2}$H$_{6}$ \cite{ricardo2012} and BF$_{3}$ \cite{ottinger1981}. An alternate method has been used to generate neutral BH molecules \cite{lester1994} from reactions between laser ablated atoms and H$_{2}$. In this process, atoms are excited to higher electronic states where collision reactions with H$_{2}$ are energetically favorable \cite{amoruso1999, lester1994}. At the H$_2$ pressures tested here, we were unable to reliably generate  BH$^+$.

\section{Conclusion and future work}

We have demonstrated the trapping and sympathetically cooling of B$^{+}$ ions by laser ablation. We confirm their presence non-destructively using the motional resonance coupling method. We have also generated various molecular ions (B$_{2}^{+}$, and B$_{3}^{+}$). Our findings provide new opportunities for experiments on a B$^{+}$ ion-based atomic clock. With the help of  theoretical predictions \cite{yangTheochem2001, yang2001}, these crystals could be used to explore the spectroscopy of the cluster ions. 

The sensitivity and resolution of our measurement method  can be enhanced by addition of a time-of-flight mass spectrometer (TOF) to our system \cite{Schowalter2012}. The TOF would provide a quantitative measure of the number and types of trapped ions.  This will be critical for understanding the trap dynamics that we have observed. 


The lifetime and sympathetic cooling of B$^+$ within Ca$^+$ crystals suggest that this system will be good for performing high-precision spectroscopy of BH$^+$ as the next steps towards  direct Doppler cooling of BH$^{+}$ ions. The report of neutral AlH from laser vaporization of LiAlH$_4$ \cite{Labazan2006} suggests that ablation of NaBH$_4$ will result in neutral BH that can then be photoionized \cite{ricardo2012}.  In this case, ionized species generated from laser ablation would be kept out of the trap using an electric field outside the trap.



\begin{acknowledgments}
This work was supported by the Army Research Office (ARO)  (W911NF-12-1-0230), the National Science Foundation (PHY-1404388), and an ARO Multi-University Research Initiative (W911NF-14-1-0378).
\end{acknowledgments}

\bibliographystyle{apsrev}

\end{document}